\documentclass[prl,aps,twocolumn,showpacs,superscriptaddress,10pt]{revtex4-1}
\usepackage[dvips]{graphicx}
\usepackage{latexsym}
\usepackage{amsmath}

\newcommand{\bpm}{\begin{pmatrix}}
\newcommand{\epm}{\end{pmatrix}}

\begin{document}


\title{Integer quantum Hall effect for bosons: A physical realization}
\author{T. Senthil}
\affiliation{Department of Physics, Massachusetts Institute of Technology,
Cambridge, MA 02139, USA}
\author{Michael Levin}
\affiliation{Condensed Matter Theory Center, Department of Physics, University of Maryland,
College Park, MD 20742, USA}

\date{\today}
\begin{abstract}
A simple physical realization of an integer quantum Hall state of interacting two
dimensional bosons is provided. This is an example of a ``symmetry-protected topological''
(SPT) phase which is a generalization of the concept of topological insulators to
systems of interacting bosons or fermions. Universal physical properties of the boson
integer quantum Hall state are described and shown to correspond to those expected from
general classifications of SPT phases.

\end{abstract}
\newcommand{\be}{\begin{equation}}
\newcommand{\ee}{\end{equation}}
\maketitle

\textsl{Introduction:}
Consider a system of two dimensional bosons with conserved particle number in the absence of
time reversal symmetry. Can such a system form a gapped phase that is qualitatively
different from a conventional Mott insulator, but has no ``intrinsic''
topological order -- i.e. no fractional excitations and a unique ground state on 
topologically non-trivial manifolds? Recent work shows that the answer is yes. In fact, 
according to the powerful cohomology classification scheme of 
Ref. \onlinecite{chencoho2011}, there are infinite number of such phases, with each phase 
labeled by an integer $n \neq 0$.

More generally, Ref. \onlinecite{chencoho2011} proposed a classification scheme for
bosonic phases with \emph{arbitrary symmetry} (time reversal, particle number
conservation, etc.) and no intrinsic topological order. These phases have been called
``symmetry-protected topological'' (SPT) phases and can be regarded as generalizations
of integer quantum Hall states and topological insulators to interacting systems of
either bosons or fermions. Loosely speaking, SPT phases are characterized by the
fact that their ground state wave functions have ``short-range entanglement'', but are
nevertheless distinct from a product state, such as a conventional Mott insulator.
More physically, SPT phases are distinguished by the presence of robust
edge modes that cannot be gapped out or localized unless the relevant symmetry
is broken. \cite{clw11,lg12}

Very recently Lu and Vishwanath\cite{luav2012} provided a beautifully simple discussion of
such symmetry-protected topological phases in two dimensions in terms of a Chern-Simons
approach and a classification of the associated $K$-matrices in the presence of symmetries.
(A similar analysis was given in Ref. \onlinecite{ls12} for the case of time reversal and 
charge conservation symmetry). Their description gives easy access to the universal
properties of such phases. For the specific case we consider here, namely bosons with a
$U(1)$ charge conservation symmetry\cite{note}, Ref. \onlinecite{luav2012} showed that the integer 
that labels these phases can be physically interpreted in terms of a quantized electric 
Hall conductivity. Specifically, the phase labeled by $n$ has an electric Hall conductivity of 
$\sigma_{xy} = 2n$ in appropriate units. Hence, these phases can be thought of as 
integer quantum Hall states for bosons. 

Here we describe a simple physical system where the simplest integer quantum Hall state
may be realized. (An alternative realization using a coupled wire construction is discussed 
in Lu and Vishwanath\cite{luav2012}). Specifically we show that a system of two component bosons 
in a strong magnetic field admits a stable integer quantum Hall phase. A natural realization is 
in terms of pseudospin-$1/2$ ``spinor" bosons of ultracold atoms in artificial gauge fields.
We analyze the basic physical properties of this  state and show
that they agree with the results expected from the general classification of Refs. \onlinecite{chencoho2011,luav2012}. 
In particular the particle number Hall conductivity is quantized to be two while the thermal Hall 
conductivity is quantized to $0$. This is related to the presence of two branches of counterpropagating chiral edge modes 
-- one which carries particle current and one which is neutral -- that are protected by the global charge $U(1)$ symmetry.  
As a bonus we show that when pseudospin $SU(2)$ symmetry is present the gapless edges cannot be gapped even if boson number conservation is explicitly broken. Thus in this situation this state may equally well be viewed as an example of an $SU(2)$ symmetric SPT state which also is predicted to occur by the classification of Ref. \onlinecite{chencoho2011}. 

\textsl{The model:}
We consider a two-component system of bosons (for instance spinor bosons or a bilayer system) 
in a strong magnetic field $B$ such that each component is at filling factor $\nu = 1$. 
Initially we assume that there is no inter-species tunneling but will relax this assumption 
later. Without inter-species tunneling, the system actually has $U(1) \times U(1)$ symmetry 
corresponding to separate conservation of the two species of bosons. The Hamiltonian is
\begin{eqnarray}
H & = & \sum_I H_I + H_{int} \\
H_I & = & \int d^2x b^\dagger_I \left(-\frac{\left(\vec \nabla - i\vec A \right)^2}{2m}
- \mu \right) b_I \\
H_{int} & = & \int d^2x d^2x' \rho_I(x) V_{IJ}(x - x') \rho_J(x')
\end{eqnarray}
Here $b_I$ is the boson annihilation operator for species $I$ where $I = 1,2$ and
$\rho_I = b^\dagger_I b_I$ is the corresponding boson density. The vector potential
$\vec A$ describes the external $B$-field. We assume that the interactions $V_{IJ}$ are
short-ranged and repulsive.

Depending on the detailed form of the interactions, a number of different states may
be realized by this system. Here we focus on a particular candidate state which
corresponds to the integer quantum Hall phase discussed above. Later we will
discuss some of the other possible competing phases.

To construct our candidate state, we use a flux attachment Chern-Simons theory. We
define new boson operators
\begin{eqnarray}
\tilde{b}_1(x) & = & e^{-i\int d^2 x' \Theta (x - x') \rho_{2}(x')} \cdot b_1(x) \\
\tilde{b}_2(x) & = & e^{-i\int d^2 x' \Theta (x - x')\rho_{1}(x')} \cdot b_2(x)
\end{eqnarray}
where $\Theta(x)$ is the angle at which the vector $\vec x$ points. This implements a flux
attachment where each boson is attached to one flux quantum of the other species.
We will call the bosons $\tilde{b}_{1,2}$ ``mutual composite bosons.''
With $\nu = 1$ for each species, we can clearly cancel the flux of the external magnetic
field in a flux smearing mean field approximation. Following the usual quantum Hall logic, an
effective Chern-Simons Landau-Ginzburg theory may be written down in terms of these mutual
composite bosons and takes the form
\begin{eqnarray}
{\cal L} & = & \sum_I {\cal L}_I + {\cal L}_{int} + {\cal L}_{CS} \nonumber \\
{\cal L}_I & = &  i\tilde{b}_I^* (\partial_0 - i A_{I 0} + i\alpha_{I 0}) \tilde{b}_I 
- \frac{|\vec \nabla \tilde{b}_I -i(\vec A_I -\vec \alpha_I) \tilde{b}_I|^2}{2m} \nonumber \\
&+& \mu |\tilde{b}_I |^2
\nonumber \\
{\cal L}_{int} & = & - V_{IJ} |\tilde{b}_I|^2 |\tilde{b}_J|^2 \nonumber \\
{\cal L}_{CS} & = & \frac{1}{4\pi} \epsilon^{\mu\nu\lambda}
\left(\alpha_{1\mu} \partial_\nu \alpha_{2\lambda} + \alpha_{2\mu}
\partial_\nu \alpha_{1\lambda}\right) \label{csterm}
\end{eqnarray}
Here we have introduced two gauge fields $\alpha_1$ and $\alpha_2$ coupled by a mutual
Chern-Simons term to implement the flux attachment. For convenience we have also introduced
external probe gauge fields $A_I$ which couple to the boson currents of species $I$.

As the mutual composite bosons see zero average flux, we can imagine a situation in which
they condense. This will lock the internal gauge field $\alpha_I$ to the probe external
field $A_I$. The effective Lagrangian for the probe gauge fields then becomes
\begin{equation}
{\cal L}_{eff} = \frac{1}{4\pi} \epsilon^{\mu\nu\lambda}
\left(A_{1\mu} \partial_\nu A_{2\lambda} +
A_{2\mu} \partial_\nu A_{1\lambda}\right)
\end{equation}
Let us now define new probe gauge fields that couple to the total charge and pseudospin
currents: $A_c = \frac{A_1 + A_2}{2}, A_s = \frac{A_1 - A_2}{2}$. In terms of these
fields,
\begin{equation}
{\cal L}_{eff} = \frac{1}{2\pi} \epsilon^{\mu\nu\lambda}
\left(A_{c\mu} \partial_\nu A_{c\lambda}
- A_{s\mu} \partial_\nu A_{s\lambda} \right)
\end{equation}
It follows that this state is incompressible and has a quantized electric Hall
conductivity of $\sigma_{xy} = +2$ in appropriate units. It is thus an integer quantum Hall state
of bosons. We can also see that this state has a pseudospin Hall conductivity of $-2$.
However, this quantity is less robust as pseudospin conservation can be broken by
inclusion of inter-species tunneling.

Several implications follow from the nonzero value for the electric Hall conductivity.
First, we can see that the above quantum Hall state belongs to a different phase
from the conventional Mott insulator (whose Hall conductivity vanishes). Second,
we conclude that the above system has robust gapless edge modes, which cannot
be gapped out unless charge conservation symmetry is broken (either explicitly
or spontaneously).

For certain purposes, it is useful to describe this state using the $K$-matrix
formalism for abelian Chern-Simons theory. Starting from the Chern-Simons term
(\ref{csterm}), it is not hard to show that the above state
corresponds to a $K$-matrix $K = \bpm 0 & 1 \\ 1 & 0 \epm$, and a charge vector
$t^T = (1,1)$. An important consequence of this identification is that the system
does not support quasiparticle excitations with fractional charge or fractional
statistics: in general such fractionalized excitations require a $K$-matrix
with with $|\det(K)| > 1$.

To obtain a model for the edge modes, we use the bulk-edge correspondence for
abelian Chern-Simons theory. According to this correspondence, the edge theory
corresponding to the above $K$-matrix is given by
\begin{equation}
{\cal L} = \frac{1}{4\pi} \left(\partial_x \phi_1 \partial_t \phi_2 +
\partial_x \phi_2 \partial_t \phi_1 - v_{IJ} \partial_x \phi_I \partial_x \phi_J \right)
\end{equation}
where $\frac{1}{2\pi} \partial_x \phi_I$ describes the density of bosons in layer $I$,
and where $v_{IJ}$ is the velocity matrix. Diagonalizing the above action, it is easy
to check that the edge contains two counterpropagating chiral modes -- one of which
carries electric charge, and one of which is electrically neutral (but carries pseudospin).
As a result of this structure, the thermal Hall conductivity vanishes, even though
the electric Hall conductivity is nonzero. We note that the vanishing of the thermal Hall conductivity
distinguishes the above integer quantum Hall state from another class of un-fractionalized
bosonic phases which require no symmetry at all -- namely the phases discussed in 
Refs. \onlinecite{kunpub} and \onlinecite{luav2012}. These phases have a non-vanishing thermal 
Hall conductivity which is a multiple of $8$ (in appropriate units).

At a microscopic level, we can understand the stability of the edge as arising from the
fact that backscattering between the two counterpropagating modes is prohibited by the
$U(1)$ charge conservation symmetry. In this sense, the edge modes are
``symmetry-protected.'' More generally, edge reconstruction may modify the above picture,
but properties such as the Hall response and overall edge stability are universal.

\textsl{Wave function:}
It is tempting but incorrect to guess that the ground state wave function for the above
bosonic quantum Hall state is simply
\begin{equation}
\label{wavefn}
\Psi\left(\{z_i, w_j\}\right) = \prod_{i, j} \left(z_i - w_j\right) \cdot
e^{- \sum_i \frac{|z_i|^2 + |w_i|^2}{4}}
\end{equation}
where $z_i, w_i$ label the complex spatial coordinates of the two species of bosons.
In the standard Halperin notation for bilayer quantum Hall states, this
is a $(001)$ state of bosons. This wave function is unstable to spontaneous phase
separation, as is readily seen using the plasma analogy: one can check that the plasma
has attractive logarithmic interactions between the pseudospin densities which implies
a spontaneous ordering of the pseudospin density, i.e. phase separation. A modified
wave function which describes a uniform boson integer quantum Hall state may be written
down as:
\begin{eqnarray}
\Psi_{mod} &=& \prod_{i< j}|z_i - z_j| \cdot \prod_{i<j} |w_i - w_j| \nonumber \\
&\cdot& \prod_{i, j}\frac{ \left(z_i - w_j\right)}{|z_i - w_j|} \cdot
e^{- \sum_i \frac{|z_i|^2 + |w_i|^2}{4}}
\label{wf1}
\end{eqnarray}
In the plasma analogy this wave function has the same effective ``Hamiltonian'' as the
$(110)$ state and hence describes a uniform density fluid with $\nu = 1$ for either species.
Indeed, a direct derivation of the wave function from the composite boson theory along the
lines of Ref. \onlinecite{zhang} yields precisely $\Psi_{mod}$.

Alternatively, we can construct a ground state wave function using a mean field flux
attachment procedure similar to the one used in Ref. [\onlinecite{wdj1993}] to construct
a spin-singlet $\nu = 2/3$ state. First, we imagine
attaching $-1$ flux quanta to each boson, thereby transforming the system into
a bilayer of composite fermions at filling $1/2 \oplus 1/2$. We then imagine that
the composite fermions form a $(111)$ state. Finally, we project onto the lowest
Landau level. The resulting wave function is:
\begin{eqnarray}
\Psi_{flux} &=& P_{LLL} \prod_{i < j} |z_i - z_j|^2 \cdot \prod_{i < j} |w_i - w_j|^2 \nonumber \\
&\cdot& \prod_{i , j} (z_i - w_j) \cdot e^{- \sum_i \frac{|z_i|^2 + |w_i|^2}{4}}
\label{wf2}
\end{eqnarray}
where $P_{LLL}$ denotes the projection onto the lowest Landau level.

An interesting feature of the two wave functions (\ref{wf1}),(\ref{wf2}), is that
they are spin singlets under the $SU(2)$ pseudospin symmetry. (One way to see this
is to note that, before projection, both wave functions can be written as a product
of the anti-analytic $(221)$ state and a fully symmetric function of $z_i, w_j$.
Given that the $(221)$ state is a spin singlet, it follows that both of the above wave
functions are also spin singlets). This enhanced symmetry means that we can equally well
regard the bosonic integer quantum Hall state as an example of an SPT phase with $SU(2)$
pseudospin symmetry, rather than $U(1)$ particle number conservation symmetry. In particular, 
explicit breaking of the global $U(1)$ symmetry by adding spin singlet pairing terms to the 
Hamiltonian will not gap out the edges so long as pseudospin $SU(2)$ symmetry is preserved. 
Thus either of these two symmetries will lead to protected gapless edge modes, due to the 
nonzero electric and pseudospin Hall conductivities, respectively. 

\textsl{Competing states:}
We now discuss some of the possible competitors to the above integer quantum Hall state.
One possibility is phase separation: the two species of bosons may clump together in
different spatial regions. Such phase separated states are particularly
natural in the limit where the inter-species interaction $V_{12}$ is large compared with the
same species interactions $V_{11}, V_{22}$. Interestingly, a fully phase separated
state has $\nu =2$ in each puddle and therefore may realize a $k=4$ non-abelian Read-Rezayi
state\cite{cwg01,cooperap08}. Another potential competitor is a decoupled state where the two
species of bosons form uncorrelated $\nu = 1$ states. Such decoupled states are likely to be 
realized in the limit where $V_{12}$ is small compared with $V_{11}, V_{22}$. This 
possibility is also potentially interesting since bosons at $\nu =1$ may form a $k=2$ 
non-abelian Read-Rezayi state which is just the familiar Moore-Read Pfaffian 
state\cite{cwg01}. A third competitor is the $\nu =2$ non-abelian spin singlet state of 
Ardonne and Schoutens\cite{asc99}. This state is a good candidate at or near the $SU(2)$ 
symmetric point where $V_{12} = V_{11} =V_{22}$.

Determining the specific circumstances under which the bosonic integer quantum Hall state
wins out over its competitors is a detailed energetics question that will not be attempted
here. We simply note that the integer quantum Hall state is a reasonable candidate
in the regime where the inter-species interaction $V_{12}$ is comparable to
the same species interactions $V_{11}, V_{22}$. We mention in passing that very recent 
numerical work\cite{gjbl12,ueda12} suggests that, in the case of delta-function repulsive 
interactions, the ground state at the $SU(2)$ symmetric point is a gapped spin singlet. Among
other possibilities, this state could be the integer quantum Hall state discussed here, or 
the non-abelian spin singlet state of Ref. \onlinecite{asc99,arrsc01}.

An obvious experimental context to seek a realization of this phase is in ultracold atoms
in artificial gauge fields. In that context, instead of a bilayer it will be simpler to
use spinor bosons and let the boson spin play the role of the bilayer index. Given
the large number of interesting competing phases in this system, it seems worthwhile
to explore experimentally the phases of spinor bosons at a total filling factor $\nu = 2$.

\textsl{Non-linear sigma model description:}
It is interesting to view the boson integer quantum Hall state from a 
different point of view. As is well known\cite{fetter} the bose condensate phase of 
two-component bosons is described by an $SU(2)$ matrix order parameter (with in general 
$O(2) \times O(2)$ anisotropy). To be specific, write the fields $b_{1,2}$ in terms of their
real and imaginary parts $b_1 = b_{1r} + ib_{1i}, b_2 = b_{2r} + i b_{2i}$, and restrict them
to the surface $|b_1|^2 + |b_2|^2 = 1$. This can be organized as an $SU(2)$
matrix $g = b_{1r} + i b_{1i} \tau^z + i b_{2r} \tau^x + i b_{2i}\tau^y$ where $\vec \tau$
are the usual Pauli matrices. It is clear that the charge $U(1)$ symmetry acts by right
multiplication by $e^{i\tau^z \chi}$ while the pseudospin $U(1)$ is generated by left
multiplication by $e^{i\tau^z \chi}$. (Full pseudospin $SU(2)$ rotation symmetry, if present, 
is realized as left multiplication by an $SU(2)$ matrix). 

It is natural to attempt a description of the phases of the two component boson system in terms 
of a quantum non-linear sigma model based on this $SU(2)$ matrix order parameter. The bose condensed 
phase of course has $\langle g \rangle \neq 0$. Disordered phases where $\langle g \rangle = 0$ 
correspond to the strong coupling limit of such a non-linear sigma model. In general, 
the effective non-linear sigma model for an $SU(2)$ matrix-valued order parameter in two space 
dimensions admits an interesting topological $\theta$ term corresponding to $\pi_3(SU(2)) = Z$. 

We now argue that the boson integer quantum Hall phase of this paper is obtained in this sigma model 
description when the $\theta$ parameter is $2\pi$ while trivial boson insulators (for instance 
an ordinary boson Mott insulator) have $\theta = 0$. To see see this, consider the $K$-matrix description 
of the boson integer quantum Hall state. Note that $K = \bpm 1 & 0 \\ 0 & 1 \epm$ means there are two bosons 
($b_1, b_2$) whose vortices see each other as $2\pi$ flux. More generally, let us consider a model of two species 
of bosons where the vortex of one picks up a phase $\theta$ when winding around the other. In a
space-time path integral picture, this means that when the two kinds of vortex world lines
link once there is an associated phase of $\theta$. Following the discussion in \cite{tsmpaf2006}, such a
model of two boson species is equivalent to an $SU(2)$ matrix non-linear sigma model with a
topological $\theta$ term and $O(2) \times O(2)$ anisotropy. When $\theta = 2\pi$ (or close to it) and in the limit where the sigma model has full
$SO(4) \sim SU(2)_R \times SU(2)_L$ symmetry, the boundary to the vacuum is described by a
$1+1$ dimensional $SU(2)$ level-1 WZW model\cite{xul12} where the $SU(2)_R$ and $SU(2)_L$ currents move in
opposite directions. Introducing the $O(2) \times O(2)$ anisotropy we recover precisely the
edge structure of the bosonic integer quantum Hall state, with the electrically charged
edge mode moving in one direction and the pseudospin edge mode moving in the opposite direction. 

The topological non-linear sigma model at $\theta = 2\pi$ also plays a crucial role 
in the cohomology classification\cite{chencoho2011}. Thus the discussion in this section provides 
a connection between the $K$-matrix description of the boson integer quantum Hall 
state and the cohomology classification.   
Recently, simulations of $O(2) \times O(2)$ models with
$\theta$ terms have appeared\cite{lesik12a,lesik12b}. It should be interesting to examine
disordered phases of these near $\theta = 2\pi$ and study their boundary to $\theta = 0$
insulators. 

\textsl{Discussion:}
In this note, we have constructed an integer quantum Hall state for bosons
with an electric Hall conductivity of $\sigma_{xy} = 2$ (in appropriate units). It is natural to 
wonder whether it is possible to construct a more ``elementary'' integer quantum Hall state -- that is, 
a state with $\sigma_{xy} =1$. We now argue that such a state is impossible if the system does 
not support fractional quasiparticle excitations. To see this, consider 
a general bosonic quantum Hall state, and imagine puncturing it at some point $z_0$ and adiabatically inserting 
$2\pi$ flux through the hole. This operation will create an excitation at $z_0$ with charge $\sigma_{xy}$. 
Let us consider the braiding statistics of these excitations. If we braid one excitation around another, 
the statistical phase follows from the Aharonov-Bohm effect: $\theta = 2\pi \sigma_{xy}$. Similarly, 
if we exchange two particles, the associated phase is $\theta/2 = \pi \sigma_{xy}$. On the other hand, 
if the state does not support fractional quasiparticles, then these excitations (like all other
quasiparticle excitations) must be bosons. We conclude that $\sigma_{xy}$ must be 
even for any bosonic quantum Hall state without fractional quasiparticle excitations.

We expect that the construction in this paper can be extended to other
symmetry-protected topological phases, such as bosonic topological insulators which are 
protected by time reversal and charge conservation symmetry, and bosonic phases which are 
protected by a discrete $Z_n$ symmetry. The boson integer quantum Hall state described 
here provides a very simple prototypical example of these phases that in addition may 
be realizable in future experiments.

\acknowledgments
We thank Xie Chen, Zheng-Xin Liu, Ashvin Vishwanath, Fa Wang and Xiao-Gang Wen for
discussions. TS was supported by NSF Grant DMR-1005434. ML was supported in part by
an Alfred P. Sloan Research Fellowship.

\appendix


\begin{thebibliography}{99}
\bibitem{note} We use the terms `boson charge' and `boson number' interchangeably through out the paper.
\end{thebibliography}
\end{document}